\documentclass[10pt, twocolumn, twoside]{IEEEtran}

\usepackage{cite}
\usepackage{enumerate}
\usepackage{graphicx}
\usepackage[cmex10]{amsmath} 
\usepackage{amssymb}
\usepackage{subfig}
\usepackage{xspace}
\usepackage[lined,linesnumbered,ruled,commentsnumbered]{algorithm2e} 
\usepackage{colonequals}
\usepackage[mathscr]{euscript}
\usepackage{fixltx2e}    
\usepackage{amsthm}
\usepackage{mathrsfs}

\usepackage{epsfig}
\usepackage{epstopdf}

\usepackage{tikz}
\usepackage{pgfplots}
\usepackage{pgfplotstable}

\usetikzlibrary{arrows,shapes,backgrounds,plotmarks,positioning,automata,calc,trees}

\pgfplotsset{compat=newest}                         
\pgfplotsset{plot coordinates/math parser=false}
\newlength\figureheight
\newlength\figurewidth

\newtheorem{theorem}{Theorem}[section]
\newtheorem{lemma}[theorem]{Lemma}

\newtheorem{example}[theorem]{Example}

\newcommand{\RZ}[1]{\mathsf{Z}_{#1}}
\newcommand{\RW}[1]{\mathsf{W}_{#1}}

\newcommand{\RCO}{\alpha^*}
\newcommand{\Card}[1]{|#1|}
\newcommand{\Set}[1]{\{#1\}}

\newcommand{\Pat}{\mathcal{P}}
\newcommand{\rv}{\mathbf{r}}
\newcommand{\rvU}{\mathbf{r}_V^{*}}  
\newcommand{\wv}{\mathbf{w}}         

\newcommand{\Vu}{f_{\alpha}}
\newcommand{\V}[1]{f_{#1}}

\newcommand{\VuHash}{f^\#_{\alpha}}
\newcommand{\VHash}[1]{f^\#_{#1}}
\newcommand{\VuHashHat}{\hat{f}^\#_{\alpha}}
\newcommand{\VHashHat}[1]{\hat{f}^\#_{#1}}

\newcommand{\DEP}{\text{dep}}      
\newcommand{\Chain}{\mathcal{C}}     
\newcommand{\DA}{\text{DA}}
\newcommand{\SFM}{\text{SFM}}

\newcommand{\RealP}{\mathbb{R}_{+}}    
\newcommand{\LEXGE}{\geq_{\text{lex}}}

\begin{document}

\title{Fairness in Communication for Omniscience}

\author{Ni~Ding\IEEEauthorrefmark{1}, Chung~Chan\IEEEauthorrefmark{3}, Qiaoqiao~Zhou\IEEEauthorrefmark{3}, Rodney~A.~Kennedy\IEEEauthorrefmark{1} and Parastoo~Sadeghi\IEEEauthorrefmark{1}
\thanks{\IEEEauthorblockA{\IEEEauthorrefmark{1}Ni Ding, Rodney A. Kennedy and Parastoo Sadeghi are with the Research School of Engineering, College of Engineering and Computer Science, the Australian National University (email: $\{$ni.ding, rodney.kennedy, parastoo.sadeghi$\}$@anu.edu.au).} The work of Ni Ding (email: ni.ding@inc.cuhk.edu.au) has been done when she was a junior research assistant with the Institute of Network Coding at Chinese University of Hong Kong from Nov 23, 2015 to Feb 6, 2016.}
\thanks{\IEEEauthorblockA{\IEEEauthorrefmark{3}Chung Chan (email: cchan@inc.cuhk.edu.hk) and Qiaoqiao Zhou (email: zq115@ie.cuhk.edu.hk) are with the Institute of Network Coding at Chinese University of Hong Kong. }}
}

\maketitle

\begin{abstract}
We consider the problem of how to fairly distribute the minimum sum-rate among the users in communication for omniscience (CO). We formulate a problem of minimizing a weighted quadratic function over a submodular base polyhedron which contains all achievable rate vectors, or transmission strategies, for CO that have the same sum-rate. By solving it, we can determine the rate vector that optimizes the Jain's fairness measure, a more commonly used fairness index than the Shapley value in communications engineering. We show that the optimizer is a lexicographically optimal (lex-optimal) base and can be determined by a decomposition algorithm (DA) that is based on submodular function minimization (SFM) algorithm and completes in strongly polynomial time. We prove that the lex-optimal minimum sum-rate strategy for CO can be determined by finding the lex-optimal base in each user subset in the fundamental partition and the complexity can be reduced accordingly.
\end{abstract}


\section{introduction}

Communication for omniscience (CO) is a problem proposed in \cite{Csiszar2004}. It is assumed that there is a group of users in the system and each of them observes a component of a discrete memoryless multiple source in private. The users can exchange their information in order to attain \textit{omniscience}, the state that each user obtains the total information in the entire multiple source in the system. A typical example is the coded cooperative data exchange (CCDE) problem \cite{Roua2010} where a group of geographically close users communicate with each other via error-free broadcast channels in order to recover a packet set.

The fundamental problem in CO and CCDE is how to achieve omniscience with minimum total transmission rate, or minimum sum-rate. The studies in \cite{ChanMMI,Ding2015ISIT} show that the minimum sum-rate can be obtained by solving an optimization problem over the partitions of user set, while the authors in \cite{CourtIT2014,MiloIT2015} propose polynomial time algorithms that determine the minimum sum-rate and a corresponding strategy based on submodular function minimization (SFM) techniques. However, the minimum sum-rate strategy is not unique in general, and the algorithms in \cite{CourtIT2014,MiloIT2015} utilizing Edmond's greedy algorithm \cite{Edmonds2003Convex} necessarily return an extremal point, or an unfair minimum sum-rate strategy.

On the other hand, fairness is an important factor in CO. For example, in CCDE, the users are considered as peers in wireless communications. A fair transmission strategy encourages them to take part in CO and helps prevent driving the battery usage of some users. Fairness has been considered in \cite{MiloDivConq2011,Ding2015ICT,Taj2011} for CCDE and in \cite{Ding2015Game} for CO. However, the methods proposed in \cite{MiloDivConq2011,Ding2015ICT} can only determine an integer valued fairest minimum sum-rate strategy in terms of Jain's fairness index \cite{Jain1984}, which may not be applied to general CO systems where the transmission rates could be fractional; the method in \cite{Taj2011} gives a fractional optimizer but relies on building a multi-layer hypergraph model which can be quite complex for large scale systems; the Shapley value proposed in \cite{Ding2015Game} incurs exponential complexity and does not coincide with the Jain's fairest solution, the commonly used fairness measure in communications engineering. We will give example to show the difference between Jain's fairest solution and Shapley value in this paper.

The main purpose of this paper is to determine the fairest solution in the minimum sum-rate strategy set for CO. We consider the problem of minimizing a weighted quadratic function on a constraint set. The constraint set contains all achievable transmission strategies that have the same sum-rate which is greater than or equal to the minimal one for CO. This problem is equivalent to determining the Jain's fairest transmission strategy when all users are assigned the same weight. We show that the constraint set of this problem is a submodular base polyhedron and the minimizer is the lexicographically optimal (lex-optimal) base. The problem can be reduced to an SFM problem, and the lex-optimal base can be determined by a decomposition algorithm (DA) in $O(|V|^2\cdot\SFM(|V|))$ time, where $|V|$ is the cardinality of the user set and $\SFM$ is the complexity of minimizing a submodular function which is strongly polynomial. We also show that the lex-optimal minimum sum-rate strategy can be determined by obtaining the lex-optimal base for each user subset in the \textit{fundamental partition} \cite{ChanMMI}, the optimal partition corresponding to the minimum sum-rate. The task of determining the lex-optimal base in a user subset is less complex than in the ground/entire user set and can be completed in parallel in distributed systems.

\section{System Model}
\label{sec:system}

Let $V=\Set{1,2,\dotsc}$ be a finite set. We assume that there are $\Card{V}>1$ users in the system. Let $\RZ{V}=(\RZ{i}:i\in V)$ be a vector of discrete random variables indexed by $V$. For each $i\in V$, user $i$ can privately observe an $n$-sequence $\RZ{i}^n$ of the random source $\RZ{i}$ that is i.i.d.\ generated according to the joint distribution $P_{\RZ{V}}$. We allow users exchange their sources directly so as to let all of them recover the source sequence $\RZ{V}^n$. Let $\rv_V=(r_i:i\in V)$ be a rate vector indexed by $V$. We call $\rv_V$ an achievable rate vector, or transmission strategy, if omniscience is possible by letting users communicate with the rates designated by $\rv_V$. Let $r$ be the function associated with $\rv_V$ such that $r(X)=\sum_{i\in X} r_i,\forall X \subseteq V$. For $X,Y \subseteq V$, let $H(\RZ{X})$ be the amount of randomness in $\RZ{X}$ measured by Shannon entropy and $H(\RZ{X}|\RZ{Y})$ be the conditional entropy of $\RZ{X}$ given $\RZ{Y}$. It is shown in \cite{Csiszar2004} that an achievable rate vector must satisfy the Slepian-Wolf constraints:
$ r(X) \geq H(\RZ{X}|\RZ{V\setminus X}), \forall X \subset V $.
Let $\alpha\in\RealP$ and
$$ \Vu(X)=\begin{cases} H(\RZ{X}|\RZ{V\setminus X}) & X \subset V \\ \alpha & X=V \end{cases}. $$
The polyhedron and base polyhedron of $\Vu$ are respectively
\begin{equation}
    \begin{aligned}
        P(\Vu,\geq)= \Set{\rv_V \mid r(X) \geq \Vu(X),\forall X \subseteq V},\\
        B(\Vu,\geq)= \Set{\rv_V \mid \rv_V \in P(\Vu,\geq), r(V)=\alpha },  \nonumber
    \end{aligned}
\end{equation}
where $B(\Vu,\geq)$ contains all achievable rate vectors that have sum-rate equal $\alpha$. $B(\Vu,\geq)=\emptyset$ means that there is no achievable rate vector that has sum-rate $\alpha$. Let $\VuHash(X)=\Vu(V)-\Vu(V \setminus X), \forall X \subseteq V$ be the dual set function of $\Vu$. We have $B(\VuHash,\leq)=B(\Vu,\geq)$ \cite{Fujishige2005}.\footnote{The majority studies on CO are based on the (intersecting) submodularity of the dual set function $\VuHash$ and its base polyhedron $B(\VuHash,\leq)$, e.g., \cite{ChanMMI,Ding2015ISIT,CourtIT2014,MiloIT2015}. } Denote $\Pi(V)$ the set of all partitions of $V$. It is shown in \cite{ChanMMI,Ding2015ISIT,Ding2015Game} that $\VuHash$ is intersecting submodular and $B(\Vu,\geq)\neq\emptyset$ if $\alpha$ is no less than the \textit{minimum sum-rate}
        \begin{equation} \label{eq:MinSumRateAsym}
            \RCO(V) = \max_{\Pat \in \Pi(V) \colon |\Pat|>1} \sum_{C \in \Pat} \frac{H(\RZ{V \setminus C}|\RZ{C})}{|\Pat|-1}.
        \end{equation}
We call the maximizer of \eqref{eq:MinSumRateAsym} the \textit{fundamental partition} and denote by $\Pat^*$. If $\alpha\geq\RCO(V)$, $B(\VuHashHat,\leq)=B(\VuHash,\leq)$ where
$$ \VuHashHat(X)=\min_{\Pat\in \Pi(X)} \sum_{C\in\Pat} \VuHash(C), \quad \forall X \subseteq V$$
is the \textit{Dilworth truncation} of $\VuHash$ and $\VuHashHat$ is submodular. Usually, $B(\Vu,\geq)$ is not a singleton when it is nonempty. So, a natural question that follows is to find a rate vector in $B(\Vu,\geq)$ that distributes the transmission load as evenly as possible among the users.

\begin{example} \label{ex:main}
Consider the user set $V=\{1,2,3\}$ and let $\RW{i}$ be an independent uniformly distributed random bit. The three users observe respectively
    \begin{align}
        \RZ{1} &= (\RW{a},\RW{b},\RW{c},\RW{d},\RW{e}),   \nonumber\\
        \RZ{2} &= (\RW{a},\RW{b},\RW{f}),   \nonumber\\
        \RZ{3} &= (\RW{c},\RW{d},\RW{f}).   \nonumber
    \end{align}
In this system, it can be shown that $\RCO(V)=3.5$ and $B(\VuHash,\leq)=B(\Vu,\geq)=\emptyset$ if $\alpha<3.5$. For $\alpha=3.5$, $B(\VHash{3.5},\leq)=B(\V{3.5},\geq)=\Set{(2.5,0.5,0.5)}$ is singleton. However, for $\alpha=4$, we have $\VHash{4}$ as
\begin{equation}
    \begin{aligned}
        &\VHash{4}(\emptyset)=0,\VHash{4}(\Set{1})=3,\VHash{4}(\Set{2})=1,\VHash{4}(\Set{3})=1, \\
        &\VHash{4}(\Set{1,2})=4,\VHash{4}(\Set{1,3})=4,\VHash{4}(\Set{2,3})=3,  \\
        &\VHash{4}(\Set{1,2,3})=4,  \nonumber
    \end{aligned}
\end{equation}
and its Dilworth truncation as
\begin{equation}
    \begin{aligned}
        &\VHashHat{4}(\emptyset)=0,\VHashHat{4}(\Set{1})=3,\VHashHat{4}(\Set{2})=1,\VHashHat{4}(\Set{3})=1, \\
        &\VHashHat{4}(\Set{1,2})=4,\VHashHat{4}(\Set{1,3})=4,\VHashHat{4}(\Set{2,3})=2,  \\
        &\VHashHat{4}(\Set{1,2,3})=4.  \nonumber
    \end{aligned}
\end{equation}
It can be shown that $B(\VHashHat{4},\leq)=B(\VHash{4},\leq)=B(\V{4},\geq)$ is a convex region, as shown in Fig.~\ref{fig:JainVsShapley}, instead of a singleton.
\end{example}

In this paper, we assume that $\alpha\geq\RCO(V)$. We will show that this assumption holds for all elements in the fundamental partition $\Pat^*$ in Section~\ref{sec:main}. Since $B(\VuHashHat,\leq)=B(\VuHash,\leq)=B(\Vu,\geq)$ when $\alpha\geq\RCO(V)$, we mainly present the results in terms of $\VuHashHat$. In Section~\ref{sec:complex}, we will show how to solve the fairness problem by the oracle calls of $\VuHash$.\footnote{The oracle takes $X \subseteq V$ as an input and outputs $\VuHash(X)$. }

\section{Motivation: Jain's Fairness vs. Shapley Value}
\label{sec:JainVsShapley}

It is shown in \cite{Ding2015Game} that a fair rate vector in $B(\VuHashHat,\leq)$ can be determined by the Shapley value $\hat{\rv}_V$ \cite{Sapley1971Convex}. Each tuple $\hat{r}_i$ of $\hat{\rv}_V$ quantifies the average marginal contribution of user $i$ in CO, and $\hat{\rv}_V$ is the gravity center of $B(\VuHashHat,\leq)$ \cite{Sapley1971Convex}. But, the Jain's fairest\footnote{Jain's fairness index of $\rv_V$ is $\frac{(\sum_{i\in V}r_i)^2}{|V|\sum_{i\in V}r_i^2}$ ranges from $\frac{1}{|V|}$ (unfairest) to $1$ (fairest) \cite{Jain1984}.} rate vector in $B(\VuHashHat,\leq)$ is the minimizer of
\begin{equation} \label{eq:Jain}
	\min\Set{\sum_{i\in V} r_i^2 \mid \rv_V\in B(\VuHashHat,\leq)}
\end{equation}
that does not usually coincide with the Shapley value.

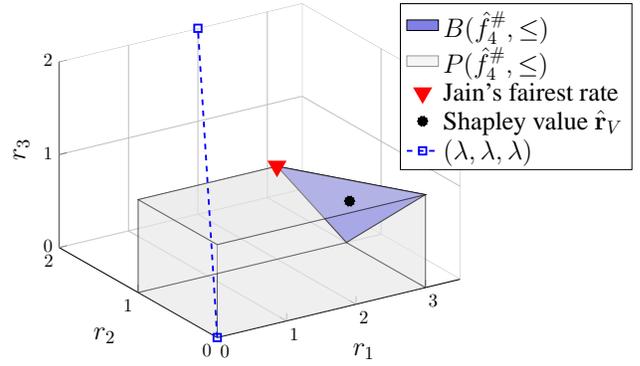
\begin{figure}[tbp]
	\centering
    \scalebox{0.7}{
%
%
%
\definecolor{mycolor1}{rgb}{0.5,0.5,0.9}%
\definecolor{mycolor2}{rgb}{1,1,0.5}%
\begin{tikzpicture}

\begin{axis}[%
width=3in,
height=2.5in,
view={-33}{30},
scale only axis,
xmin=0,
xmax=3.5,
xlabel={\Large $r_1$},
xmajorgrids,
ymin=0,
ymax=2,
ylabel={\Large $r_2$},
ymajorgrids,
zmin=0,
zmax=2,
zlabel={\Large $r_3$},
zmajorgrids,
axis x line*=bottom,
axis y line*=left,
axis z line*=left,
legend style={at={(0.85,1)},anchor=north west,draw=black,fill=white,legend cell align=left}
]

%

\addplot3[area legend,solid,fill=mycolor1,draw=black]
table[row sep=crcr]{
x y z\\
2 1 1 \\
3 0 1 \\
3 1 0 \\
2 1 1 \\
};
\addlegendentry{\Large $B(\VHashHat{4},\leq)$};

\addplot3[area legend,solid,fill=white!90!black,opacity=4.000000e-01,draw=black]
table[row sep=crcr]{
x y z\\
0 0 0 \\
3 0 0 \\
3 1 0 \\
0 1 0 \\
0 0 0 \\
};
\addlegendentry{\Large $P(\VHashHat{4},\leq)$};

\addplot3[solid,fill=white!90!black,opacity=4.000000e-01,draw=black,forget plot]
table[row sep=crcr]{
x y z\\
0 0 0 \\
0 1 0 \\
0 1 1 \\
0 0 1 \\
0 0 0 \\
};

\addplot3[solid,fill=white!90!black,opacity=4.000000e-01,draw=black,forget plot]
table[row sep=crcr]{
x y z\\
0 0 0 \\
3 0 0 \\
3 0 1 \\
0 0 1 \\
0 0 0 \\
};

\addplot3[solid,fill=white!90!black,opacity=4.000000e-01,draw=black,forget plot]
table[row sep=crcr]{
x y z\\
3 0 0 \\
3 1 0 \\
3 0 1 \\
3 0 0 \\
};

\addplot3[solid,fill=white!90!black,opacity=4.000000e-01,draw=black,forget plot]
table[row sep=crcr]{
x y z\\
0 1 0 \\
0 1 1 \\
2 1 1 \\
3 1 0 \\
0 1 0 \\
};

\addplot3[solid,fill=white!90!black,opacity=5.000000e-01,draw=black,forget plot]
table[row sep=crcr]{
x y z\\
0 0 1 \\
3 0 1 \\
2 1 1 \\
0 1 1 \\
0 0 1 \\
};

\addplot3 [
color=red,
line width=4.0pt,
only marks,
mark=triangle,
mark options={solid,,rotate=180}]
table[row sep=crcr] {
2 1 1\\
};
\addlegendentry{\Large Jain's fairest rate};

\addplot3 [
color=black,
line width=5.0pt,
only marks,
mark=star,
mark options={solid}]
table[row sep=crcr] {
2.66666666666667 0.666666666666667 0.666666666666667\\
};
\addlegendentry{\Large Shapley value $\hat{\rv}_V$};

\addplot3 [
color=blue,
dashed,
line width=1.0pt,
mark=square,
mark options={solid}]
table[row sep=crcr] {
0 0 0\\
2 2 2\\
};
\addlegendentry{\Large $(\lambda,\lambda,\lambda)$};

%

\end{axis}
\end{tikzpicture}%
}
	\caption{In the CO system in Example~\ref{ex:main}, the Shapley value $\hat{\rv}_V=(\frac{8}{3},\frac{2}{3},\frac{2}{3})$ is the gravity center of $B(\VHashHat{4},\leq)$. The Jain's fairest rate vector is $\rvU=(2,1,1)\neq\hat{\rv}_V$. Instead, $\rvU=(2,1,1)=\arg\max\Set{r(V) \mid \rv_V \in P(\VHashHat{4},\leq), \rv_V \leq (2,2,2)}$. }
	\label{fig:JainVsShapley}
\end{figure}

\begin{example} \label{ex:JainVsShapley}
    In the system in Example~\ref{ex:main}, in $B(\VHashHat{4},\leq)$, the Shapley value is $\hat{r}_V=(\frac{8}{3},\frac{2}{3},\frac{2}{3})$, while the Jain's fairest rate vector is $(2,1,1)$. See Fig.~\ref{fig:JainVsShapley}.
\end{example}

It can be easily seen based on the definition of Shapley value in \cite{Sapley1971Convex} that $\hat{\rv}_V$ is usually biased so that the user has more information on $\RZ{V}$ transmits more. For example, in the system in Example~\ref{ex:main}, we have $H(\RZ{1})$ greater than $H(\RZ{2})$ and $H(\RZ{3})$. The three vertex points of $B(\VHashHat{4},\leq)$, $(2,1,1)$, $(3,0,1)$ and $(3,1,0)$, all have $r_1$ greater than the other entries, and so does the gravity center $\hat{\rv}_V$. However, in communications or network engineering, we want to distribute the rates as evenly as possible without considering the users' prior knowledge about the sources. So, the ideal rate vector is in the form of $\lambda \cdot \chi_V$, where $\chi_X$ is the characteristic or incidence vector of $X \subseteq V$. If such a rate vector does not belong to $B(\VuHashHat,\leq)$, we need to at least find one that is as close as possible to it, e.g. the rate vector $(2,1,1)$ in Fig.~\ref{fig:JainVsShapley}. In this sense, the minimizer of \eqref{eq:Jain} provides a better solution than Shapley value, and it is worth discussing on how to solve problem \eqref{eq:Jain} efficiently.

\section{Main Results}

Let $\wv_V$ be a positive weight vector. We consider a problem that is more general than \eqref{eq:Jain}:
\begin{equation} \label{eq:LexOpt}
	 \min\Set{\sum_{i\in V} \frac{r_i^2}{w_i} \mid \rv_V\in B(\VuHashHat,\leq)}.
\end{equation}
The objective function can be considered as a weighted Jain's fairness measure in that \eqref{eq:LexOpt} reduces to \eqref{eq:Jain} when $\wv_V=\chi_V=(1,\dotsc,1)$. The main purpose of this section is to show that the fairest minimum sum-rate strategy for CO can be determined based on the fundamental partition $\Pat^*$. We start this section by presenting some existing results on the minimization problem \eqref{eq:LexOpt} in Section~\ref{sec:LexOpt}, where we show that the minimizer of \eqref{eq:LexOpt} is a lexicographically optimal (lex-optimal) base in $B(\VuHashHat,\leq)$. Based on these results, we present our main result in Section~\ref{sec:main}: The lex-optimal minimum sum-rate strategy is a direct merge of the the lex-optimal bases of the user subsets in the fundamental partition $\Pat^*$.

\subsection{Lex-optimal Base}
\label{sec:LexOpt}

A set $X$ is called $\rv_V$-tight if $r(X)=\VuHashHat(X)$.\footnote{An $\rv_V$-tight set $X$ means rate saturation in entries $r_i$ for all $i \in X$, i.e., due to the constraint $r(X)\leq\VuHashHat(X)$, for $\xi>0$, $\rv_V+\xi \cdot \chi_i \notin P(\VuHashHat,\leq),\forall i \in X$. The example in Section~\ref{sec:path} shows how tight set is related to the minimizer $\rvU$ of \eqref{eq:LexOpt}. } Let $\DEP(\rv_V,i)$ be the smallest $\rv_V$-tight set that includes $i$. For $\rv_V\in B(\VuHashHat,\leq)$, $\DEP(\rv_V,i)$ can be expressed  as \cite{Fujishige2005}
$$ \DEP(\rv_V,i)=\Set{ j \mid \exists{\xi>0} \colon \rv_V+\xi(\chi_i-\chi_j)\in B(\VuHashHat,\leq) }, $$
where $\xi(\chi_i-\chi_j)$ is called \textit{elementary transformation}.
It is shown in \cite{Fujishige2005} that the local optimality with respect to the direction $\chi_i-\chi_j$ implies the global optimality: $\rvU$ is the minimizer of \eqref{eq:LexOpt} if
$ \sum_{i\in V} \frac{(r_i^*)^2}{w_i}\leq \sum_{i\in V} \frac{r_i^2}{w_i}, \forall \rv_V=\rvU+\xi(\chi_i-\chi_j)\in B(\VuHashHat,\leq)$. The condition is equivalent to \eqref{eq:MainCond} in the lemma below.

\begin{lemma}[{\cite[Theorem 8.1]{Fujishige2005}}] \label{lemma:main}
    $\rvU$ is the minimizer of \eqref{eq:LexOpt} iff
    \begin{equation} \label{eq:MainCond}
        \frac{r_i^*}{w_i} \geq \frac{r_j^*}{w_j}, \quad \forall j\in\DEP(\rvU,i)\setminus\Set{i}.
    \end{equation}
\end{lemma}

Let $T_{\wv_V}(\rv_V)=(\frac{r_{\sigma(1)}}{w_{\sigma(1)}},\dotsc,\frac{r_{\sigma(|V|)}}{w_{\sigma(|V|)}})$ where $\sigma(i)$ is an ordering of user indices such that $\frac{r_{\sigma(1)}}{w_{\sigma(1)}} \leq \dotsc \leq \frac{r_{\sigma(|V|)}}{w_{\sigma(|V|)}}$, e.g., $T_{(4,2,1)}((2,3,1))=(0.5,1,1.5)$ where $\sigma(1)=1$, $\sigma(2)=3$ and $\sigma(3)=2$. It is shown in \cite{Fujishige2009PP} that, for $\rvU$ and any other $\rv_V\in B(\VuHashHat,\leq)$, we have $\frac{r_{\sigma(i)}^*}{w_{\sigma(i)}}=\frac{r_{\sigma(i)}}{w_{\sigma(i)}}, \forall i>i'$ and $\frac{r_{\sigma(i')}^*}{w_{\sigma(i')}}>\frac{r_{\sigma(i')}}{w_{\sigma(i')}}$ for some $i'$. It is called $T_{\wv_V}(\rvU)$ lexicographically dominates $T_{\wv_V}(\rv_V)$ and denoted by $T_{\wv_V}(\rvU) \LEXGE T_{\wv_V}(\rv_V)$. So, $\rvU$ is called the \textit{lex-optimal base} in $B(\VuHashHat,\leq)$ w.r.t. $\wv_V$.

\begin{example} \label{ex:lexopt}
  In the system in Example~\ref{ex:main}, for $\wv_V=(4,2,1)$, we have lex-optimal base $\rvU=(2.4,1,0.6)$, for which $\DEP(\rvU,1)=\Set{1,2,3}$, $\DEP(\rvU,2)=\Set{2}$, $\DEP(\rvU,3)=\Set{1,2,3}$ and Lemma~\ref{lemma:main} holds. For example, for $i=1, j=2$, we have $j\in\DEP(\rvU,i)\setminus\Set{i}$, $\frac{r_i^*}{w_i}=0.6 \geq \frac{r_j^*}{w_j}=0.5$. Also, $T_{\wv_V}(\rvU)=(0.5,0.6,0.6)\LEXGE T_{\wv_V}(\rv_V)$ for any other $\rv_V\in B(\VHashHat{4},\leq)$. For example, for $\rv_V=(2.2,1,0.8)\in B(\VHashHat{4},\leq)$, we have $T_{\wv_V}(\rv_V)=(0.5,0.55,0.8)$ and $T_{\wv_V}(\rvU)\LEXGE T_{\wv_V}(\rv_V)$ because the $1$st entries of $T_{\wv_V}(\rv_V)$ and $T_{\wv_V}(\rvU)$ are equal while the $2$nd entry of $T_{\wv_V}(\rvU)$ is greater than that of $T_{\wv_V}(\rv_V)$. If we change $\wv_V$ to $(1,1,1)$, we have $\rvU=(2,1,1)$. It also can be shown that Lemma~\ref{lemma:main} holds and $(2,1,1)$ is the lex-optimal base in $B(\VHashHat{4},\leq)$.
\end{example}

\subsection{Lex-optimal minimum sum-rate strategy}
\label{sec:main}

Let the fundamental partition be $\Pat^*=\Set{C_1,\dotsc,C_{|\Pat^*|}}$ where $\emptyset \neq C_m \subset V,\forall m\in\Set{1,\dotsc,|\Pat^*|}$. We have \cite{ChanMMI}
$$\RCO(V)=\sum_{C_m\in\Pat^*} \VHash{\RCO(V)}(C_m)=\sum_{C_m\in\Pat^*} \VHashHat{\RCO(V)}(C_m). $$
Also, for each $C_m\in\Pat^*$, we have \cite{MiloDivConq2011}
$$\VHash{\RCO(V)}(C_m) = \VHashHat{\RCO(V)}(C_m) \geq \RCO(C_m). $$ Alternatively speaking, in order to achieve omniscience with the minimum sum-rate $\RCO(V)$, each user subset $C_m$ in the fundamental partition $\Pat^*$ must transmit exactly $\VHashHat{\RCO(V)}(C_m)$ times which is greater than or equal to the minimum sum-rate for achieving the omniscience in $C_m$ itself. Let $h_{C_m}(X)=\VHashHat{\RCO(V)}(X),\forall X \subseteq C_m$. For $C_m,C_{m'}\in\Pat^*$ such that $C_m \neq C_{m'}$, denote $\rv_{C_m} \oplus \rv_{C_{m'}} = (r_i \colon i \in C_m \sqcup C_{m'})$ the direct merging of $\rv_{C_m}$ and $\rv_{C_{m'}}$. For example, for $\rv_{C_m}=(r_1,r_4,r_5)=(2,3,4)$ and $\rv_{C_{m'}}=(r_2,r_3)=(4,1)$, $\rv_{C_m} \oplus \rv_{C_{m'}} = (r_1,\dotsc,r_5) = (2,4,1,3,4) $. The following theorem states that the lex-optimal minimum sum-rate strategy w.r.t. $\wv_V$ is a direct merge of the lex-optimal base in each user subset $C_m$ w.r.t. $\wv_{C_m}$ in the fundamental partition $\Pat^*$.

\begin{theorem}  \label{theo:minsumratefair}
    The lex-optimal base in $B(\VHashHat{\RCO(V)})$  w.r.t. $\wv_V$ is $\rvU = \rv_{C_1}^* \oplus \dotsc \oplus \rv_{C_{|\Pat^*|}}^*$, where $\Pat^*$ is the fundamental partition and $\rv_{C_m}^*$ is the lex-optimal base in $B(h_{C_m},\leq)$ w.r.t. $\wv_{C_m}$.
\end{theorem}
\begin{IEEEproof}
    For any $\rv_V \in B(\VHashHat{\RCO(V)},\leq)$, $r(C_m)=\VHashHat{\RCO(V)}(C_m)=h_{C_m}(C_m),\forall C_m \in \Pat^*$, i.e., $C_m$ is $\rv_V$-tight for all user subset $C_m$ in the fundamental partition $\Pat^*$. Let $\rv'_V= \rv_{C_1}^* \oplus \dotsc \oplus \rv_{C_{|\Pat^*|}}^*$ where $\rv_{C_m}^*$ is the lex-optimal base of $B(h_{C_m},\leq)$ w.r.t. $\wv_{C_m}$. Then, $\rv'_V \in B(\VHashHat{\RCO(V)},\leq)$ necessarily \cite[Lemma 3.1]{Fujishige2005}. Also, for all $i \in C_m \in \Pat^*$, $\DEP(\rv'_V,i) \subseteq C_m$ and $\frac{r'_i}{w_i} \geq \frac{r'_j}{w_j},\forall j\in\DEP(\rvU,i)\setminus\Set{i}$, which is equivalent to condition~\eqref{eq:MainCond} in Lemma~\ref{lemma:main}. Therefore, $\rv'_V$ is the lex-optimal base w.r.t. $\wv_V$.
\end{IEEEproof}

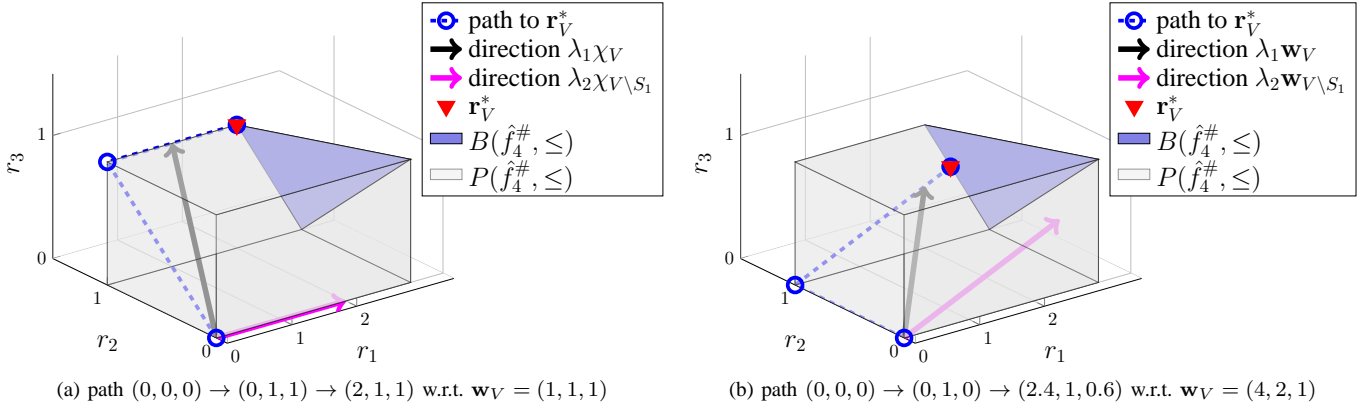
\begin{figure*}[t]
	\centering
    \subfloat[path $(0,0,0)\rightarrow(0,1,1)\rightarrow(2,1,1)$ w.r.t. $\wv_V=(1,1,1)$]{\scalebox{0.7}{
%
%
%
\definecolor{mycolor1}{rgb}{1,0,1}%
\definecolor{mycolor2}{rgb}{0.5,0.5,0.9}%
\begin{tikzpicture}

\begin{axis}[%
width=3in,
height=2.5in,
view={322.5}{30},
scale only axis,
xmin=0,
xmax=3.5,
xlabel={\Large $r_1$},
xtick={0, 1, 2},
xmajorgrids,
ymin=-0.1,
ymax=1.5,
ylabel={\Large $r_2$},
ytick={-1,0, 1, 2},
ymajorgrids,
zmin=0,
zmax=1.5,
zlabel={\Large $r_3$},
ztick={0, 1, 2},
zmajorgrids,
axis x line*=bottom,
axis y line*=left,
axis z line*=left,
legend style={at={(0.92,1.02)},anchor=north west,draw=black,fill=white,legend cell align=left}
]

%

\addplot3 [
color=blue,
dashed,
line width=2.0pt,
mark size=4.0pt,
mark=o,
mark options={solid}]
table[row sep=crcr] {
0 0 0\\
0 1 1\\
2 1 1\\
};
\addlegendentry{\Large path to $\rvU$};

\addplot3 [
->,
line width=3.0pt,]
table[row sep=crcr] {
0 0 0\\
1 1 1\\
};
\addlegendentry{\Large direction $\lambda_1\chi_{V}$};

\addplot3 [
->,
color=mycolor1,
line width=3.0pt,]
table[row sep=crcr] {
0 0 0\\
2 0 0\\
};
\addlegendentry{\Large direction $\lambda_2\chi_{V \setminus S_1}$};

\addplot3 [
color=red,
line width=4.0pt,
only marks,
mark=triangle,
mark options={solid,,rotate=180}]
table[row sep=crcr] {
2 1 1\\
};
\addlegendentry{\Large $\rvU$};



\addplot3[area legend,solid,fill=mycolor2,draw=black]
table[row sep=crcr]{
x y z\\
2 1 1 \\
3 0 1 \\
3 1 0 \\
};

\addlegendentry{\Large $B(\VHashHat{4},\leq)$};

\addplot3[area legend,solid,fill=white!90!black,opacity=4.000000e-01,draw=black]
table[row sep=crcr]{
x y z\\
0 0 0 \\
3 0 0 \\
3 1 0 \\
0 1 0 \\
0 0 0 \\
};

\addlegendentry{\Large $P(\VHashHat{4},\leq)$};

\addplot3[solid,fill=white!90!black,opacity=4.000000e-01,draw=black,forget plot]
table[row sep=crcr]{
x y z\\
0 0 0 \\
0 1 0 \\
0 1 1 \\
0 0 1 \\
0 0 0 \\
};

\addplot3[solid,fill=white!90!black,opacity=4.000000e-01,draw=black,forget plot]
table[row sep=crcr]{
x y z\\
0 0 0 \\
3 0 0 \\
3 0 1 \\
0 0 1 \\
0 0 0 \\
};

\addplot3[solid,fill=white!90!black,opacity=4.000000e-01,draw=black,forget plot]
table[row sep=crcr]{
x y z\\
0 0 0 \\
3 0 0 \\
3 0 1 \\
0 0 1 \\
0 0 0 \\
};

\addplot3[solid,fill=white!90!black,opacity=4.000000e-01,draw=black,forget plot]
table[row sep=crcr]{
x y z\\
3 0 0 \\
3 1 0 \\
3 0 1 \\
3 0 0 \\
};

\addplot3[solid,fill=white!90!black,opacity=4.000000e-01,draw=black,forget plot]
table[row sep=crcr]{
x y z\\
0 1 0 \\
0 1 1 \\
2 1 1 \\
3 1 0 \\
0 1 0 \\
};

\addplot3[solid,fill=white!90!black,opacity=5.000000e-01,draw=black,forget plot]
table[row sep=crcr]{
x y z\\
0 0 1 \\
3 0 1 \\
2 1 1 \\
0 1 1 \\
0 0 1 \\
};

\end{axis}
\end{tikzpicture}%
}} \quad
    \subfloat[path $(0,0,0)\rightarrow(0,1,0)\rightarrow(2.4,1,0.6)$ w.r.t. $\wv_V=(4,2,1)$]{\scalebox{0.7}{
%
%
%
\definecolor{mycolor1}{rgb}{1,0,1}%
\definecolor{mycolor2}{rgb}{0.5,0.5,0.9}%
\begin{tikzpicture}

\begin{axis}[%
width=3in,
height=2.5in,
view={322.5}{30},
scale only axis,
xmin=0,
xmax=3.5,
xlabel={\Large $r_1$},
xtick={0, 1, 2},
xmajorgrids,
ymin=-0.1,
ymax=1.5,
ylabel={\Large $r_2$},
ytick={-1,0, 1, 2},
ymajorgrids,
zmin=0,
zmax=1.5,
zlabel={\Large $r_3$},
ztick={0, 1, 2},
zmajorgrids,
axis x line*=bottom,
axis y line*=left,
axis z line*=left,
legend style={at={(0.92,1.02)},anchor=north west,draw=black,fill=white,legend cell align=left}
]

%

\addplot3 [
color=blue,
dashed,
line width=2.0pt,
mark size=4.0pt,
mark=o,
mark options={solid}]
table[row sep=crcr] {
0 0 0\\
0 1 0\\
2.4 1 0.6\\
};
\addlegendentry{\Large path to $\rvU$};

\addplot3 [
->,
line width=3.0pt,]
table[row sep=crcr] {
0 0 0\\
2 1 0.5\\
};
\addlegendentry{\Large direction $\lambda_1\wv_{V}$};

\addplot3 [
->,
color=mycolor1,
line width=3.0pt,]
table[row sep=crcr] {
0 0 0\\
2.4 0 0.6\\
};
\addlegendentry{\Large direction $\lambda_2\wv_{V \setminus S_1}$};



\addplot3 [
color=red,
line width=4.0pt,
only marks,
mark=triangle,
mark options={solid,,rotate=180}]
table[row sep=crcr] {
2.4 1 0.6\\
};
\addlegendentry{\Large $\rvU$};

\addplot3[area legend,solid,fill=mycolor2,draw=black]
table[row sep=crcr]{
x y z\\
2 1 1 \\
3 0 1 \\
3 1 0 \\
};

\addlegendentry{\Large $B(\VHashHat{4},\leq)$};

\addplot3[area legend,solid,fill=white!90!black,opacity=4.000000e-01,draw=black]
table[row sep=crcr]{
x y z\\
0 0 0 \\
3 0 0 \\
3 1 0 \\
0 1 0 \\
0 0 0 \\
};

\addlegendentry{\Large $P(\VHashHat{4},\leq)$};

\addplot3[solid,fill=white!90!black,opacity=4.000000e-01,draw=black,forget plot]
table[row sep=crcr]{
x y z\\
0 0 0 \\
0 1 0 \\
0 1 1 \\
0 0 1 \\
0 0 0 \\
};

\addplot3[solid,fill=white!90!black,opacity=4.000000e-01,draw=black,forget plot]
table[row sep=crcr]{
x y z\\
0 0 0 \\
3 0 0 \\
3 0 1 \\
0 0 1 \\
0 0 0 \\
};

\addplot3[solid,fill=white!90!black,opacity=4.000000e-01,draw=black,forget plot]
table[row sep=crcr]{
x y z\\
0 0 0 \\
3 0 0 \\
3 0 1 \\
0 0 1 \\
0 0 0 \\
};

\addplot3[solid,fill=white!90!black,opacity=4.000000e-01,draw=black,forget plot]
table[row sep=crcr]{
x y z\\
3 0 0 \\
3 1 0 \\
3 0 1 \\
3 0 0 \\
};

\addplot3[solid,fill=white!90!black,opacity=4.000000e-01,draw=black,forget plot]
table[row sep=crcr]{
x y z\\
0 1 0 \\
0 1 1 \\
2 1 1 \\
3 1 0 \\
0 1 0 \\
};

\addplot3[solid,fill=white!90!black,opacity=5.000000e-01,draw=black,forget plot]
table[row sep=crcr]{
x y z\\
0 0 1 \\
3 0 1 \\
2 1 1 \\
0 1 1 \\
0 0 1 \\
};

\end{axis}
\end{tikzpicture}%
}}
	\caption{The path towards the lex-optimal $\rvU$ w.r.t. $\wv_V$ in the system in Example~\ref{ex:main}.}
    \label{fig:Paths}
\end{figure*}

\begin{example} \label{ex:minsumratefair}
    Let the user set be $V=\{1,2,3,4\}$. The four users observe respectively
    \begin{align}
        \RZ{1} &= (\RW{c},\RW{d},\RW{f},\RW{g},\RW{h}),   \nonumber\\
        \RZ{2} &= (\RW{a},\RW{d},\RW{g},\RW{h}),   \nonumber\\
        \RZ{3} &= (\RW{c},\RW{d},\RW{e},\RW{f},\RW{g},\RW{h}),   \nonumber\\
        \RZ{4} &= (\RW{a},\RW{b},\RW{f}).   \nonumber
    \end{align}
    In this system, we have minimum sum-rate for CO $\RCO(V)=6$ and the fundamental partition $\Pat^*=\Set{\Set{1,2,3},\Set{4}}$. For $C_1=\Set{1,2,3}$, we have $h_{C_1}(X)=\VHashHat{6}(X), \forall X \subseteq C_1$ and $B(h_{C_1},\leq)\neq\emptyset$ because $h_{C_1}(C_1)=5>\RCO(C_1)=4$. Consider determining the lex-optimal base in $B(\VHashHat{6},\leq)$ w.r.t. $\wv_V=\chi_V=(1,1,1,1)$ by Theorem~\ref{theo:minsumratefair}. The lex-optimal base w.r.t. $\wv_{C_1}=(1,1,1)$ in $B(h_{C_1},\leq)$ is $\rv_{C_1}^*=(r_1^*,r_2^*,r_3^*)=(\frac{5}{3},\frac{5}{3},\frac{5}{3})$. For $C_2=
    \Set{4}$, $\rv_{C_2}^*=r_4^*=1$. The direct merging of $\rv_{C_1}$ and $\rv_{C_2}$ is $\rv_{C_1}^*\oplus\rv_{C_2}^*=(\frac{5}{3},\frac{5}{3},\frac{5}{3},1)$. It can be shown that $\rv_{C_1}^*\oplus\rv_{C_2}^*$ is the lex-optimal base w.r.t. $\wv_V=(1,1,1,1)$ in $B(\VHashHat{6},\leq)$, which is also the Jain's fairest minimum sum-rate strategy.
\end{example}

\section{Path towards $\rvU$}
\label{sec:path}

The problem unsolved in Theorem~\ref{theo:minsumratefair} is how to determine the lex-optimal base in each user subset. In this section, we use the existing results in \cite{Fujishige2009PP} to show that there exists a piecewise linear path towards the lex-optimal base, which can be found by a decomposition algorithm in the existing literature.

Let the distinct values of $\frac{r_i^*}{w_i}$ of the lex-optimal base $\rvU$ be $\lambda_1,\dotsc,\lambda_N$ such that $0=\lambda_0<\lambda_1<\dotsc<\lambda_N$. Denote
$S_0=\emptyset$ and $ S_n=\Set{i \mid \frac{r_i^*}{w_i}\leq\lambda_n, i \in V }$. We have all $S_n$ form a chain \cite{Fujishige2009PP}
$$ \Chain_S : \emptyset = S_0 \subset S_1 \subset \dotsc \subset S_N=V$$
such that $r_i^*=\lambda_n w_i$, for all $i \in S_n \setminus S_{n-1}$. Since $\DEP(\rvU,i)\subseteq S_n, \forall i \in S_n$, $r^*(S_n)=\VuHashHat(S_n)$. It means that we cannot increase the value of $r_i^*$ for all $ i \in S_n $ because otherwise the condition $r^*(S_n) \leq \VuHashHat(S_n)$ in the polyhedron $P(\VuHashHat,\leq)$ is breached. Alternatively speaking, $\rvU \wedge \lambda_n$ is the maximizer of $\max\Set{ r(V) \mid \rv_V\in P(\VuHashHat,\leq), \rv_V \leq \lambda_n \cdot \wv_V }$ \cite{Fujishige2009PP},
where $\rvU \wedge \lambda_n \cdot \wv_V=(\min\Set{r_i^*,\lambda_n \cdot w_i} \colon i \in V)$ is the componentwise minimization of $\rvU$ and $\lambda_n \cdot \wv_V$. It is clear that a piecewise linear path in $P(\VuHashHat,\leq)$ towards the lex-optimal base $\rvU$ can be determined by considering the problem
\begin{equation} \label{eq:max}
    \max\Set{ r(V) \mid \rv_V\in P(\VuHashHat,\leq), \rv_V \leq \lambda \cdot \wv_V },
\end{equation}
where $\lambda\in[0,+\infty)$. This path also determines the values of all $S_n$ and $\lambda_n$.

\begin{example} \label{ex:path}
    We explain the path $ (0,0,0) \rightarrow (0,1,1) \rightarrow (2,1,1) $ towards the lex-optimal base $\rvU=(2,1,1)$ w.r.t. $\wv_V=(1,1,1)$ in Fig.~\ref{fig:Paths}(a) by using \eqref{eq:max} as follows. We initiate $\rvU=(0,0,0)$ and the direction $\lambda \cdot \wv_V = \lambda \cdot \chi_V = (\lambda,\lambda,\lambda)$. We increase $\lambda$ from $0$ to move along the direction $\chi_V$ until at $\lambda=1$ we reach a boundary of $P(\VHashHat{4},\leq)$ that corresponds to the inequality $r(\Set{2,3}) \leq \VHashHat{4}(\Set{2,3})=2$. Here, the boundary point $(1,1,1)$ is necessarily the maximizer of $\max\Set{ r(V) \mid \rv_V\in P(\VHashHat{4},\leq), \rv_V \leq \lambda_1 \cdot \wv_V = (1,1,1) }$. We cannot increase $\lambda$ any more along the direction $\lambda \cdot \chi_V$ since otherwise we will be out of $P(\VHashHat{4},\leq)$ so that the resulting vector does not belong to $B(\VHashHat{4},\leq)$. For the set $\Set{2,3}$ that is associated with the boundary, we know that we can at least allocate $\VHashHat{4}(\Set{2,3})=2$ rates evenly to users $2$ and $3$. Therefore, by letting $\lambda_1=1$ and $S_1=\Set{2,3}$, we assign rates $r_2^*=\lambda_1 w_2=1$ and $r_3^*=\lambda_1 w_3=1$ so that the rate $\rvU$ is updated to $(0,1,1)$. By knowing that we cannot increase any further along any dimension in $S_1$,\footnote{It means that $S_1$ is $\rvU$-tight, i.e., $r^*(S_1)=\VHashHat{4}(S_1)$. } we set the direction $\lambda \cdot \chi_{V \setminus S_1} = (\lambda,0,0)$ and continue to increase the value of $\lambda$ from $\rvU=(0,1,1)$. The purpose is to see if we can distribute the remaining rates $\VHashHat{4}(V)-\VHashHat{4}(S_1)=2$ evenly among the users in $V \setminus S_1$. At $\lambda=2$, we reach the second boundary that is set by the constraint $r(\Set{1,2,3}) \leq \VHashHat{4}(\Set{1,2,3})=4$ in $P(\VHashHat{4},\leq)$. We set $\lambda_2=2$ and $S_2=\Set{1,2,3}$. For $S_2\setminus S_1=\Set{3}$, we assign $r_3^*=\lambda_2 w_3=2$. $\rvU$ is updated to $(2,1,1)$. Here, the second boundary point $(2,1,1)$ is the maximizer of $\max\Set{ r(V) \mid \rv_V\in P(\VHashHat{4},\leq), \rv_V \leq \lambda_2 \cdot \wv_V = (2,2,2) }$. Now, we have $V \setminus S_2=\emptyset$, which means all $\VHashHat{4}(V)=4$ rates have been allocated. We get $\rvU=(2,1,1)$ as the lex-optimal rate vector w.r.t. $\wv=(1,1,1)$. In the same way, we can show the path $(0,0,0)\rightarrow(0,1,0)\rightarrow(2.4,1,0.6)$ towards $\rvU=(2.4,1,0.6)$ w.r.t. $\wv_V=(4,2,1)$ in Fig.~\ref{fig:Paths}(b). Based on this path, we have $\lambda_1=0.5$ corresponds to $S_1=\Set{2}$ and $\lambda_2=0.6$ corresponds to $S_2=\Set{1,2,3}$.
\end{example}

	\begin{algorithm} [t]
	\label{algo:DAPP}
	\small
	\SetAlgoLined
    \SetKwInOut{Input}{input}\SetKwInOut{Output}{output}
	\SetKwRepeat{Repeat}{repeat}{until}
    \SetKwIF{If}{ElseIf}{Else}{if}{then}{else if}{else}{endif}
    \SetKw{Return}{return}
    \Input{$\underline{S},\overline{S}$ in $\Chain_S$}
	\Output{the set that contains $\underline{S}$, $\overline{S}$ and all $S_n$ in $\Chain_S$ such that $\underline{S} \subset S_n \subset \overline{S}$  }
	\BlankLine
	\Begin{
        $\lambda=\frac{\VuHashHat(\overline{S}) - \VuHashHat(\underline{S}) }{w(\overline{S} \setminus \underline{S})}$\;
        get $S$ as the maximal minimizer of $$\min \Set{\VuHashHat(X)-\lambda w(X) \colon X \subseteq V};$$
        \lIf{$S=\overline{S}$}{
                return $(\underline{S},\overline{S})$
            }\lElse{
                return $\DA(\underline{S},S)\cup\DA(S,\overline{S})$
            }
        }	
	\caption{decomposition algorithm (DA) \cite{Nagano2013}}
	\end{algorithm}

Due to the min-max theorem \cite{Edmonds2003Convex}
\begin{multline} \label{eq:minmax}
    \max\Set{ r(V) \mid \rv_V\in P(\VuHashHat,\leq), \rv_V \leq \lambda \cdot \wv_V }  = \\
    \min\Set{\VuHashHat(X) + \lambda w(V \setminus X) \mid X \subseteq V},
\end{multline}
problem \eqref{eq:max} reduces to $\min\Set{\VuHashHat(X)-\lambda w(X) \mid X \subseteq V}$, which is an SFM problem and can be efficiently solved by the DA algorithm in Algorithm 1 that is proposed in \cite{Nagano2013}.

\begin{example}
    In $B(\VHashHat{4},\leq)$ in Example~\ref{ex:main}, consider the problem of finding the lex-optimal base $\rvU$ w.r.t. $\wv_V=(4,2,1)$ by the DA algorithm. We call $\DA(\emptyset, V)$, i.e., $\underline{S}=\emptyset$ and $\overline{S}=V$. We get $\lambda=\frac{4}{7}$ and $S=\Set{2}$. Since $S \neq \overline{S}=V$, $\DA(\emptyset,\Set{2})$ is called where we get $\lambda=0.5$ and $S=\Set{2}=\overline{S}$. It returns $\Set{\emptyset,\Set{2}}$. On the other hand, in the call of $\DA(\Set{2},V)$, we get $\lambda=0.6$ and $S=V=\overline{S}$. It returns $\Set{\Set{2},V}$. We then get the chain $C_S \colon \emptyset \subset \Set{2} \subset \Set{1,2,3}=V$ corresponding to $\lambda_0=0$, $\lambda_1=0.5$ and $\lambda_2=0.6$. By calling DA algorithm to determine the lex-optimal base $\rvU$ w.r.t. $\wv_V=(1,1,1)$, we get the chain $C_S \colon \emptyset \subset \Set{2,3} \subset \Set{1,2,3}=V$ corresponding to $\lambda_0=0$, $\lambda_1=1$ and $\lambda_2=2$. The results are consistent with those in Example~\ref{ex:path}.
\end{example}

\section{Complexity}
\label{sec:complex}

In step 2 in Algorithm 1, the value of the Dilworth truncation $\VuHashHat$ at certain $X \subseteq V$ can be obtained by oracle calls of $\VuHash$ in $O(|V|\cdot\SFM(|V|))$ time, where $\SFM$ is the complexity of minimizing a submodular function. In step 3, the maximal minimizer of $\min\Set{\VuHashHat(X)-\lambda w(X) \mid X \subseteq V}$ can be obtained by considering the problem of $\min\Set{\VuHash(X)-\lambda w(X) \mid X \subseteq V \colon i \in V} $. For each $i\in V$, it completes in $O(|V|\cdot\SFM(|V|))$ time by oracle calls of $\VuHash$. To determine all $S_n$ in chain $\Chain_S$, the DA algorithm is called for no greater than $2|V|-1$ times \cite{Nagano2012Lex}. Therefore, the complexity of determining $\rvU$ by oracle calls of $\VuHash$ is $O(|V|^2\cdot\SFM(|V|))$, which is strongly polynomial as compared to the exponentially growing complexity in obtaining the Shapley value $\hat{\rv}_V$ \cite{Sapley1971Convex}.

Consider the complexity of determining the lex-optimal minimum sum-rate strategy by Theorem~\ref{theo:minsumratefair}. The fundamental partition $\Pat^*$ can be determined in $O(|V|^2\cdot\SFM(|V|))$ time \cite{MinAveCost}. After obtaining $\Pat^*$, we can determine the lex-optimal base $\rv_{C_m}^*$ for each $C_m\in\Pat^*$ and get the lex-optimal minimum sum-rate strategy $\rvU$. Since $|C_m|<|V|$, $O(|C_m|^2\cdot\SFM(|C_m|))$, the complexity of finding $\rv_{C_m}^*$, is much lower than that of finding $\rv_{V}^*$. In addition, $\rv_{C_m}^*$ for all $C_m\in\Pat^*$ can be found in parallel in distributed systems. For example, in Example~\ref{ex:minsumratefair} where $\Pat^*=\Set{\Set{1,2,3},\Set{4}}$, we can allow a user delegate in $C_1=\Set{1,2,3}$, say, user 1, and user $4$ in $C_2=\Set{4}$ to determine the lex-optimal base $\rv_{C_1}^*$ and $\rv_{C_2}^*$, respectively.

\section{Conclusion}
We considered the problem of minimizing a weighted quadratic function over the set of all achievable rate vectors with the same sum-rate for CO. The objective function was a generalization of the Jain's fairness index. We showed that if the constraint set was nonempty the optimizer was a lex-optimal base that could be searched by a DA algorithm in strongly polynomial time. We also showed that the lex-optimal minimum sum-rate strategy for CO could be determined by directly merging the lex-optimal bases for all user subset in the fundamental partition, which could be completed in a decentralized manner with lower complexity.

\bibliographystyle{ieeetr}
\bibliography{LexOpt}

\end{document}